\documentclass[prb,twocolumn,showpacs,amsmath,amssymb,preprintnumbers,superscriptaddress]{revtex4}
\usepackage{dcolumn}
\usepackage{bm}
\usepackage{graphicx}
\usepackage{color}

\begin{document}

\title{Evolution of structure, magnetism and electronic transport in doped pyrochlore iridate Y$_2$Ir$_{2-x}$Ru$_{x}$O$_7$}

\author{Harish Kumar}\affiliation{School of Physical Sciences, Jawaharlal Nehru University, New Delhi - 110067, India.}
\author{R. S. Dhaka}\affiliation{Department of Physics, Indian Institute of Technology, Hauz Khaz, New Delhi - 110016, India.}
\author{A. K. Pramanik}\email{akpramanik@mail.jnu.ac.in}\affiliation{School of Physical Sciences, Jawaharlal Nehru University, New Delhi - 110067, India.}

\begin{abstract}
The interplay between spin-orbit coupling (SOC) and electron correlation ($U$) is considered for many exotic phenomena in iridium oxides. We have investigated the evolution of structural, magnetic and electronic properties in pyrochlore iridate Y$_2$Ir$_{2-x}$Ru$_{x}$O$_7$ where the substitution of Ru has been aimed to tune this interplay. The Ru substitution does not introduce any structural phase transition, however, we do observe an evolution of lattice parameters with the doping level $x$. X-ray photoemission spectroscopy (XPS) study indicates Ru adopts charge state of Ru$^{4+}$ and replaces the Ir$^{4+}$ accordingly. Magnetization data reveal both the onset of magnetic irreversibility and the magnetic moment decreases with progressive substitution of Ru. These materials show non-equilibrium low temperature magnetic state as revealed by magnetic relaxation data. Interestingly, we find magnetic relaxation rate increases with substitution of Ru. The electrical resistivity shows an insulating behavior in whole temperature range, however, resistivity decreases with substitution of Ru. Nature of electronic conduction has been found to follow power-law behavior for all the materials.
\end{abstract}


\maketitle
\section {Introduction}
The interplay between spin-orbit coupling (SOC) and electron correlation ($U$) has recently been predicted to give rise many exotic topological phases in iridium oxides.\cite{pesin, william} The 5$d$ character of Ir renders a comparable energy scale of SOC and $U$ which promotes interesting magnetic and electronic behavior. The iridates with pyrochlore structure (A$_2$Ir$_2$O$_7$ where A  rare earth elements) are  particularly interesting due to its geometrical structure where the interpenetrating corner-shared tetradehra of A and Ir cations introduces magnetic frustration which often does not allow the system for magnetic ordering down to low temperature. However, inclusion of Dzyaloshinskii-Moriya interaction (DMI), which is very evident in systems with reasonable SOC, along with the original exchange interactions has been shown to give rise magnetic transition to long-range ordered state at low temperature in otherwise frustrated disordered systems.\cite{elhajal}

Among the pyrochlore iridates, Y$_2$Ir$_2$O$_7$ is of particular interest. The non-magnetic nature of Y$^{3+}$ avoids possible $f$-$d$ interactions,\cite{chen} hence the physical properties are mostly determined by Ir lattice. In that sense, this material offers an ideal system to study the influence of combined SOC and $U$ interactions. A suitable substitution at Ir-site in this material will effectively tune the both SOC and $U$ parameters, hence the properties of the material will modify accordingly. The Y$_2$Ir$_2$O$_7$ exhibits a magnetic irreversibility as characterized by an onset of bifurcation between zero field cooled (ZFC) and field cooled (FC) magnetization around 160 K.\cite{shapiro,disseler,taira,fukazawa,soda,zhu,harish} The low temperature magnetic state, however, remains controversial. While the neutron diffraction and inelastic measurements\cite{shapiro} show no sign of magnetic long-range transition, the muon spin relaxation and rotation investigations,\cite{disseler} on the other hand, have shown long-range type magnetic ordering at low temperature. Further, another study has shown an existence of weak ferromagnetism along with large antiferromagnetic background in Y$_2$Ir$_2$O$_7$.\cite{zhu} A recent study has shown nonequilibrium magnetic phase in Y$_2$Ir$_2$O$_7$, showing reasonable magnetic relaxation at low temperature.\cite{harish} On the theoretical side, calculation involving LDA + $U$ method has predicted a Weyl-type semimetal phase with all-in/all-out (AIAO) magnetic order in Y$_2$Ir$_2$O$_7$.\cite{wan} Similarly, other calculations employing LDA + DMFT and LSDA + $U$ methods have shown a transition to AIAO type antiferromagnetic (AFM) state with increasing $U$.\cite{shinaoka,Ishii} Interestingly, the critical value of $U$ for such magnetic transition has not been found consistently in all these studies and it appears that obtained results much depend on the adopted methods. Although, photoemission spectroscopy measurements estimate a strong $U$ ($\sim$ 4 eV) in Y$_2$Ir$_2$O$_7$ but its direct correlation with magnetic behavior is not well understood.\cite{kalo}

In present work, we investigate the effect of interplay between SOC and $U$ in pyrochlore iridates which has been realized by substituting magnetic Ru ion in Y$_2$Ir$_2$O$_7$. The Ru (4$d^4$) has relatively high $U$ and low SOC compared to Ir (5$d^5$) which would tune the both these parameters accordingly. Moreover, this substitution will amount to hole doping in the system which would modify the electronic properties across the Fermi level. On the other hand, matching ionic radii of Ir and Ru presumably will not introduce any major structural modification in this series. The structural distortions i.e., trigonal distortion has, however, been shown to have significant influence on the magnetic and electronic properties in the family of pyrochlore iridates.\cite{yang}

We have studied the evolution of structural, magnetic and electronic properties in doped Y$_2$Ir$_{2-x}$Ru$_{x}$O$_7$. The system retains its original structure with Ru substitution, but we find minor changes in local structural parameters. X-ray photoemission spectroscopy measurements show both Ir and Ru ions adopt charge state of +4 and changes the ionic concentration of Ir accordingly. Both the magnetic irreversibility temperature (T$_{irr}$) and the magnetic moment at low temperature decreases with Ru substitution. The parent compound Y$_2$Ir$_2$O$_7$ as well as its Ru doped analogues show nonequilibrium magnetic state at low temperature where the magnetic relaxation rate increases with Ru. While the samples remains insulating up to 40\% of Ru doping but the resistivity decreases and the mode of charge conduction follows power-law behavior for all the samples.      

\section {Experimental Details}
Polycrystalline samples of Y$_2$Ir$_{2-x}$Ru$_{x}$O$_7$ with $x$ = 0.0, 0.1, 0.2 and 0.4 have been prepared using standard solid state method. Details of sample preparation and characterization are given in Ref. \onlinecite{harish}. The ingredient powder materials Y$_2$O$_3$, RuO$_2$ and IrO$_2$ with phase purity 99.99\% and above (Sigma-Aldrich) are taken in stoichiometric ratio and are ground well. Then, powders are pelletized and heated in air at 1000$^o$C for 96 hours, at 1100$^o$C for 96 hours and at 1160$^o$C for 252 hours, respectively with intermediate grindings. The materials have been characterized by powder x-ray diffraction (XRD) using a Rigaku made diffractometer with Cu$K_\alpha$ radiation and by x-ray photoemission spectroscopy (XPS). XRD data have been collected in the range of 2$\theta$ = 10 - 90$^o$ at a step of $\Delta 2\theta$ = 0.02$^o$ The collected XRD data have been analyzed using Reitveld refinement program (FULLPROF) by Young \textit{et al}.\cite{young} A commercial electron energy analyzer (PHOIBOS 150 from Specs GmbH, Germany) and a non-monochromatic Al$K_\alpha$ x-ray source ($h\nu$ =  1486.6 eV) have been used to perform the XPS measurements with the base pressure in the range of 10$^{-10}$ mbar. XPS data have been analyzed using CasaXPS software. DC Magnetization data have been collected using a vibrating sample magnetometer (PPMS, Quantum Design) and Electrical transport properties have been measured using a home-built insert fitted with Oxford superconducting magnet.

\begin{figure}
	\centering
		\includegraphics[width=8cm]{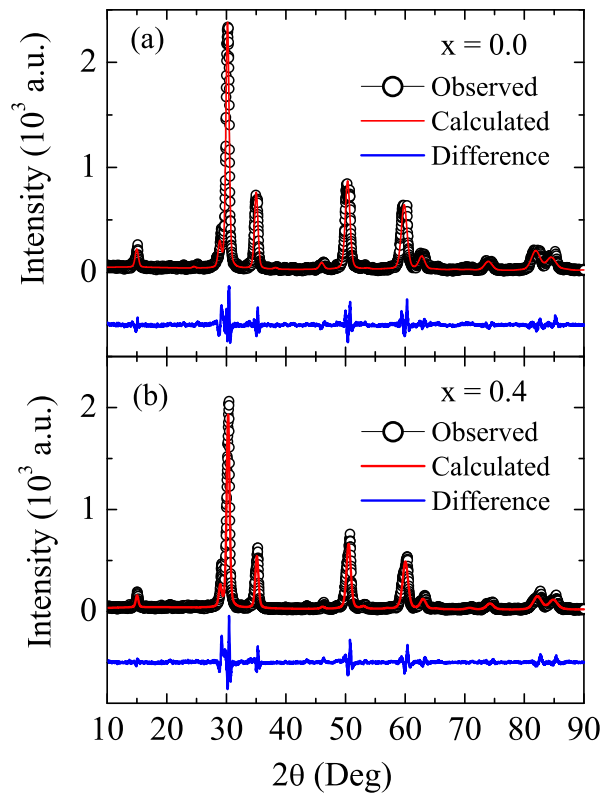}
	\caption{Room temperature XRD pattern along with Rietveld analysis have been shown for representative (a) $x$ = 0.0 and (b) $x$ = 0.4 composition for Y$_2$Ir$_{2-x}$Ru$_{x}$O$_7$ series.}
	\label{fig:Fig1}
\end{figure}

\section{Results and Discussions}
\subsection {Structural analysis}
Fig. 1 shows representative XRD pattern along with Rietveld analysis for extreme two members of present series Y$_2$Ir$_{2-x}$Ru$_{x}$O$_7$ i.e., with $x$ = 0.0 and 0.4. The XRD pattern for parent Y$_2$Ir$_2$O$_7$ match well with previous studies.\cite{shapiro,disseler,zhu} The Reitveld analysis of XRD data shows Y$_2$Ir$_2$O$_7$ crystallizes in cubic phase with \textit{Fd$\bar{3}$m} symmetry (Fig. 1a). Fig. 1b shows XRD pattern and its analysis for the compound with $x$ = 0.4. The figure shows substitution of Ru$^{4+}$ for Ir$^{4+}$ in Y$_2$Ir$_2$O$_7$ does not introduce major changes in crystal structure which is also expected from their matching ionic radii (Ir$^{4+}$ = 0.625 \AA and Ru$^{4+}$ = 0.62 \AA). Rietveld analysis, however, shows minor changes in structural parameters in this series as shown in Fig. 2. We obtain lattice constant $a$ for Y$_2$Ir$_2$O$_7$ is 10.244(1) \AA which decreases monotonically upon substitution of Ru. This changes in $a$ is not though substantial as we calculate change $\Delta a$ is about -0.23\% over the series.
 
\begin{figure}
	\centering
		\includegraphics[width=8cm]{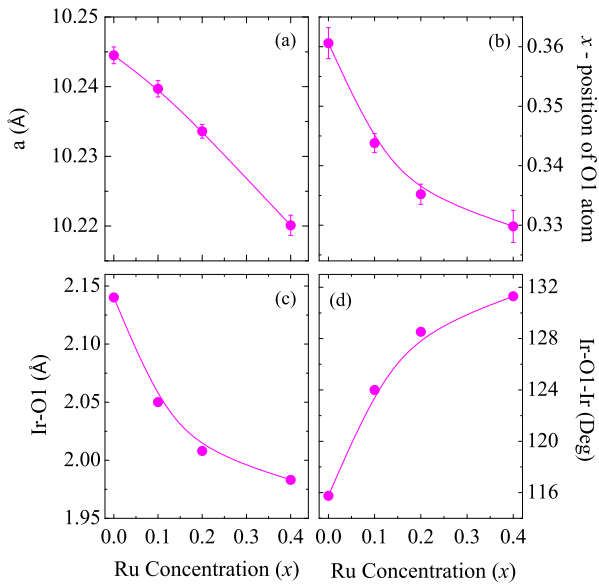}
	\caption{(a) Lattice constant $a$, (b) $x$- position of O1 atom, (c) Ir-O1 bond length and (d) Ir-O1-Ir bond angel are shown as a function of Ru substitution for Y$_2$Ir$_{2-x}$Ru$_{x}$O$_7$ series. Lines are guide to eyes.}
	\label{fig:Fig2}
\end{figure}

The structural organization of IrO$_6$ octahedra places a vital role in pyrochlore system to decide its physical properties. Fig. 2b shows position of oxygen atom attached to IrO$_6$ octahedra as represented by `$x$-position of O1 atom'. In IrO$_6$ octahedra the oxygen ions are at equal distance from central Ir ion, however, the position of oxygen can vary contributing distortion to octahedra. For instance, for $x$ = 0.3125 octahedra's are perfect and Ir ions are under ideal cubic field. Deviation of $x$ from this value generates distortion to octahedra and gives trigonal crystal field. Fig. 2b shows $x$ value for this series. For Y$_2$Ir$_2$O$_7$, the $x$ value has been found to be 0.36 which implies IrO$_6$ octahedra's are distorted and compressed. Substitution of Ru decreases the $x$ value leading towards perfect octahedra, hence trigonal crystal field is reduced. Similarly, Figs. 2c and 2d show Ir-O bond length and Ir-O1-Ir bond angle, respectively related to IrO$_6$ octahedra for present series. With substitution of Ru, bond length decreases while bond angel increases which is consistent with behavior of $a$ in Fig. 2a. This implies that distortion in IrO$_6$ octahedra is reduced and the overlap between Ir(5\textit{d})/Ru(4\textit{d}) and O(2\textit{p}) orbitals is increased.

\begin{figure*}
	\centering
		\includegraphics[width=16cm]{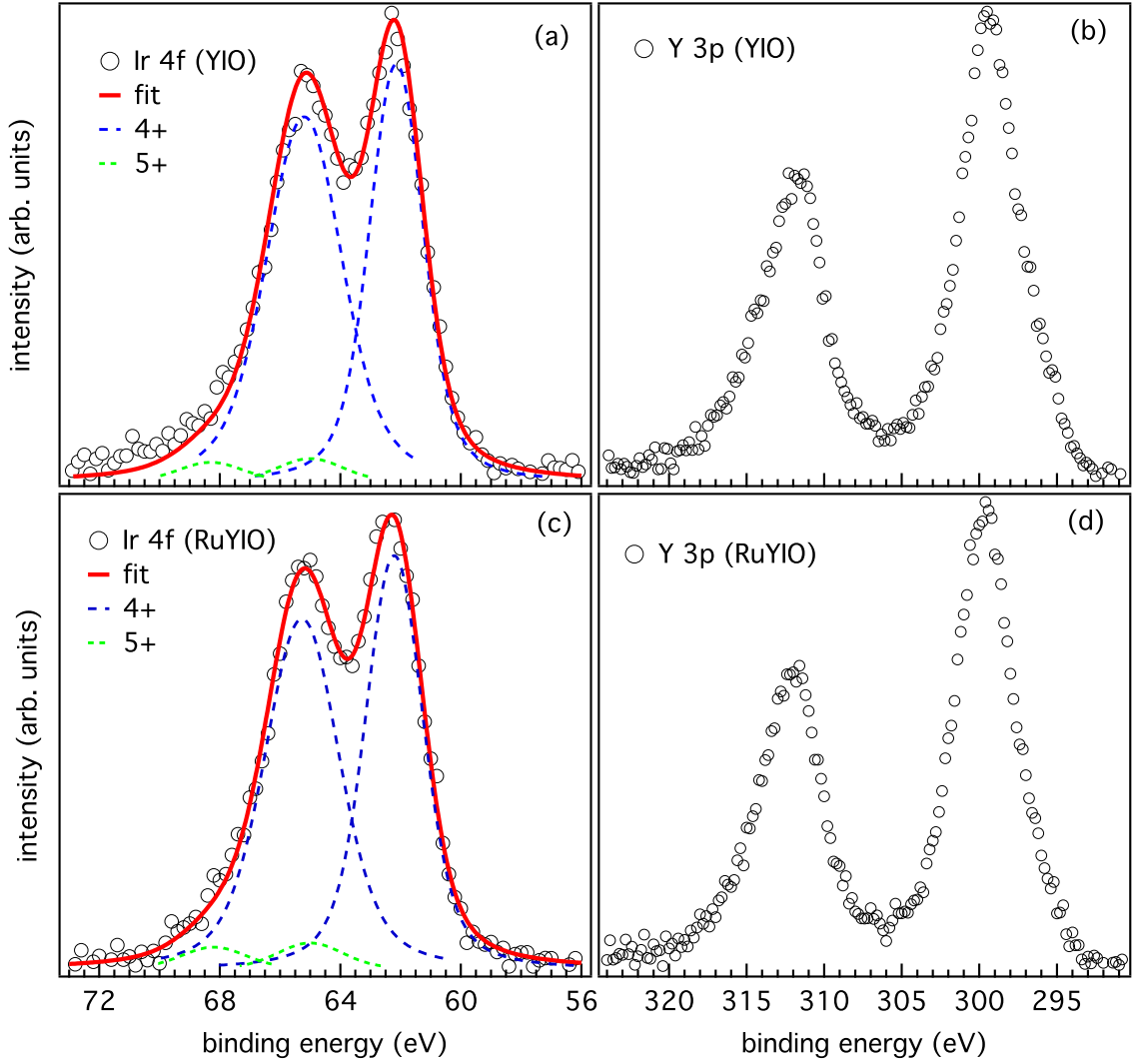}
	\caption{XPS spectra related to (a) Ir-4$f$ and (b) Y-3$p$ level are shown for parent Y$_2$Ir$_2$O$_7$ (YIO) material. Open black circles are the XPS data. The red solid line in (a) is the fitted envelope taking contributions of Ir$^{4+}$ and Ir$^{5+}$ components which are individually plotted in dashed blue and green colors, respectively. (c) and (d) show the same XPS data and related fittings for doped Y$_2$Ir$_{1.6}$Ru$_{0.4}$O$_7$ (RuYIO) material.}
	\label{fig:Fig3}
\end{figure*}

\subsection {X-ray photoemission spectroscopy study}
Understanding the charge state of transition metal is very important in this class of materials which largely governs the electronic and magnetic properties. For this purpose we have carried out XPS measurements on parent as well as Ru doped samples where data have been shown in Fig. 3. The Ir-4$f$ spectra have been shown in Figs. 3a and 3c for $x$ = 0 parent material Y$_2$Ir$_2$O$_7$ (YIO) and $x$ = 0.4 doped material Y$_2$Ir$_{1.6}$Ru$_{0.4}$O$_7$ (RuYIO), respectively. Similarly, Figs. 3b and 3d show the Y-3$p$ spectra for the same YIO and RuYIO materials, respectively. We observe that Y-3$p_{3/2}$ and Y-3$p_{1/2}$ peaks at binding energy (BE) around 299.5 and 312 eV with the spin-orbit splitting of $\sim$12.5 eV, which are in agreement with the literature.\cite{xps} Moreover, there is no measurable change in the BE of Y-3$p$ core-level between two samples. We have analyzed the Ir 4$f$ core-level spectra in more detail using standard CasaXPS software package, as shown in Figs. 3a and 3c. The continuous red lines in Figs. 3a and 3c represent overall fitting (envelope) of Ir-4$f$ data taking individual contributions of both Ir$^{4+}$ (dashed blue line) and Ir$^{5+}$ (dashed red line). For Y$_2$Ir$_2$O$_7$, our detail fitting shows majority of Ir is in Ir$^{4+}$ ionic state while there is some contribution of Ir$^{5+}$ ionic state has been observed in both the samples. For Ir$^{4+}$, spin-orbit split two peaks i.e., 4$f_{7/2}$ and 4$f_{5/2}$ electronic states appear around 62 and 65 eV of binding energies, respectively which are shown as dashed blue lines in Figs. 3a and 3c. Similarly, spectral component of Ir$^{5+}$ state appear around 65 and 68 eV (dashed green line) for 4$f_{7/2}$ and 4$f_{5/2}$, respectively. It is evident in figures that intensity of Ir$^{5+}$ component is much weaker compared to Ir$^{4+}$ component.

More detailed analysis reveal that for parent material the amount of Ir$^{4+}$ and Ir$^{5+}$ ions are about 94.6(5) and 5.4(5)\%, respectively. For the doped material Y$_2$Ir$_{1.6}$Ru$_{0.4}$O$_7$, we find that content of Ir$^{5+}$ does not change significantly but that of Ir$^{4+}$ changes in proportion with Ru substitution showing the content of Ir$^{4+}$ and Ir$^{5+}$ to be around 74.9(5) and 5.1(5)\%, respectively. The analysis shows content of Ir$^{5+}$ remains almost unchanged with Ru substitution within error bar, however, that of Ir$^{4+}$ decreases by $\sim$ 19.5\% which agrees with the nominal concentration of 20\% Ru in this material. The presence of Ir$^{5+}$ at minimal level in parent Y$_2$Ir$_2$O$_7$ material has also been reported in other study using XPS measurements.\cite{zhu} The most likely cause for the simultaneous presence of Ir$^{5+}$ would be due to excess oxygen in samples. Nonetheless, XPS data analysis shows Ru is in Ru$^{4+}$ state (not shown) and replaces Ir$^{4+}$ ions which is rather expected considering both they have almost matching charge state and ionic radii.

\begin{figure}
	\centering
		\includegraphics[width=8cm]{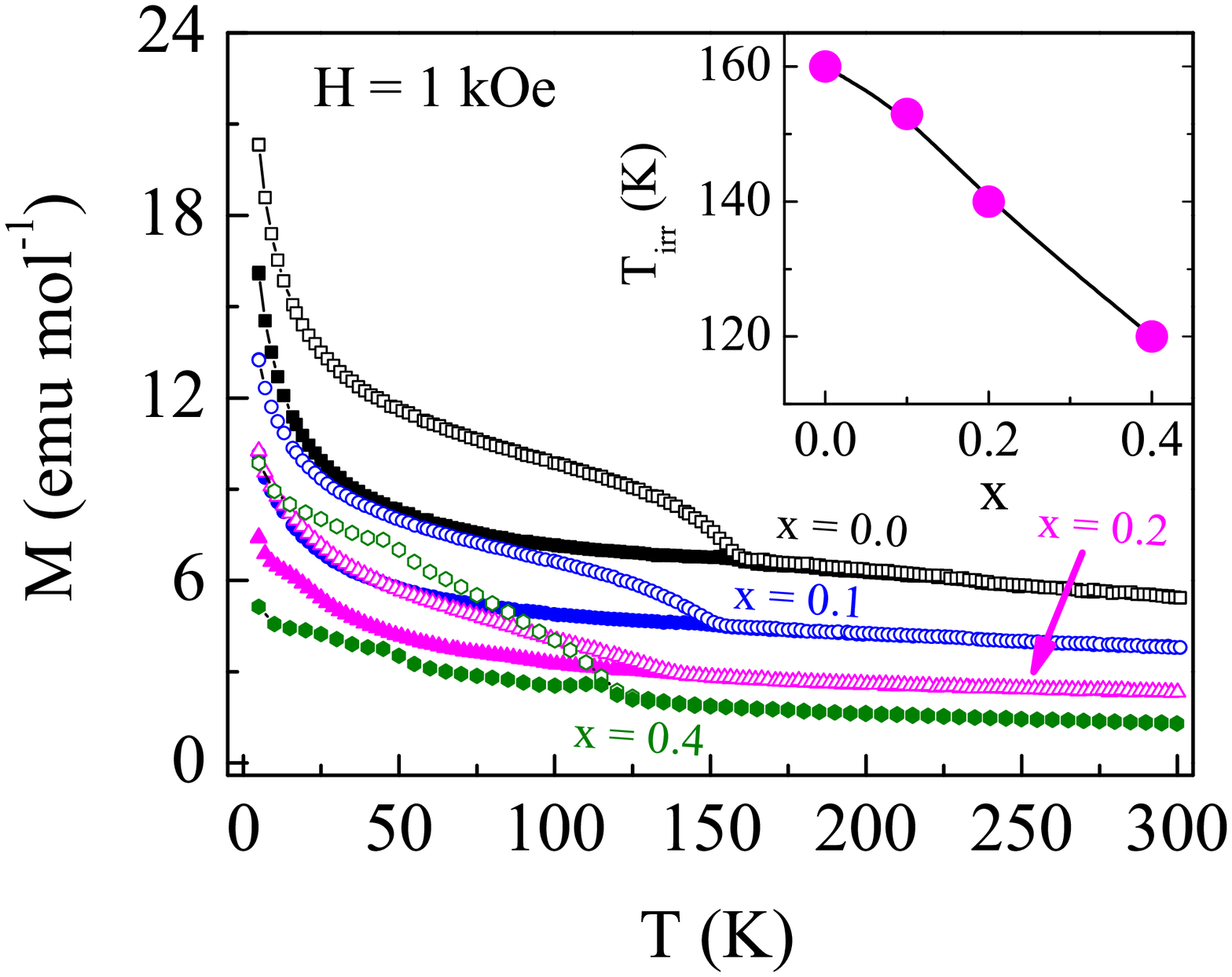}
	\caption{Temperature dependent magnetization data measured in 1 kOe under ZFC and FC protocol have been shown for Y$_2$Ir$_{2-x}$Ru$_x$O$_7$ series. Filled and unfilled symbols represent ZFC and FC data, respectively. Inset shows variation of irreversibility temperature T$_{irr}$ with Ru substitution for the same series.}
	\label{fig:Fig4}
\end{figure}

\subsection{Magnetization study}
Fig. 4 shows magnetization ($M$) data as a function of temperature ($T$) measured in 1000 Oe magnetic field following zero field cooled (ZFC) and field cooled (FC) protocol for present series. The undoped parent compound shows a clear magnetic irreversibility around temperature T$_{irr}$ = 160 K, below which a large bifurcation between the ZFC and FC magnetization is observed. It can be noted that $M_{ZFC}$ does not exhibit any prominent cusp/peak around T$_{irr}$. The nature of low temperature magnetic state in Y$_2$Ir$_2$O$_7$ is debated as neutron powder diffraction \cite{shapiro} and muon spin rotation experiments \cite{disseler} have shown contradicting results in favor of long-range magnetic order below T$_{irr}$. A recent study has shown nonequilibrium low temperature phase in Y$_2$Ir$_2$O$_7$.\cite{harish} An existence of weak ferromagnetism in association with large antiferromagnetic background has been shown by another recent study.\cite{zhu} As evident in figure, with Ru substitution the bifurcation temperature T$_{irr}$ between $M_{FC}$ and $M_{ZFC}$ shifts to lower temperature. Furthermore, ZFC magnetization shows lower value as $x$ increases. The inset of Fig. 4 presents composition dependent T$_{irr}$ showing $T_{irr}$ decreases almost linearly with Ru content. Fig. 5 shows inverse magnetic susceptibility ($\chi^{-1} = (M/H)^{-1}$) as deduced from Fig. 4 with temperature for this series. The $\chi^{-1}(T)$ shows linear behavior in high temperature regime as evident in Fig. 5a for $x$ = 0.0, 0.1 and in Fig. 5b for $x$ = 0.2, 0.4 materials. The data have been fitted in range of 200 - 300 K with modified Curie-Weiss law;

\begin{eqnarray}
	\chi = \chi_0 + \frac{C}{T - \theta_P}
\end{eqnarray}

where $\chi_0$ is temperature independent magnetic susceptibility, $C$ is the Curie constant and $\theta_P$ is Curie temperature. The effective paramagnetic moment $\mu_{eff}$ has been calculated from $C$. Using the obtained values, frustration parameter $f$ has been calculated from ratio $\left|\theta_P\right|$/$T_{irr}$ for the whole series. All the parameters have been shown in Table I. For Y$_2$Ir$_2$O$_7$, we obtain a high value of $\theta_P$ = -331 K. Both sign as well as magnitude of $\theta_P$ indicate a nonferromagnetic type of magnetic interaction with reasonable strength. With introduction of Ru the $\theta_P$ value decreases, yet it remains negative with $x$ up to 0.4. This implies that nature of magnetic interaction remains nonferromagnetic type but the interaction is weakened as we substitute Ru in system. The calculated frustration parameter $f$ suggests reasonable frustration in Y$_2$Ir$_2$O$_7$. Pyrochlore compounds in general have in-built frustration due to their geometrical arrangement which often lead to exotic magnetic ground states such as, spin-glass \cite{gingras,yoshii}, spin-liquid \cite{gardner1,nakatsuji}, spin-ice \cite{bramwell,fukazawa1}, etc. For example, $f$ values for well-known frustrated pyrochlore oxides such as, Y$_2$Ru$_2$O$_7$ and Y$_2$Mo$_2$O$_7$ are about 16 and 8.8, respectively which exhibit spin-glass behavior.\cite{kmiec,Gardner} It is interesting that with Ru substitution the $f$ parameter decreases which suggests frustration level decreases over the series.

The isothermal magnetization as a function of magnetic field ($H$) measured at 5 K are shown in Fig. 6a. The figure shows $M(H)$ plot for $x$ = 0.0 compound is not linear and the magnetization increases continuously without sign of saturation till magnetic field 70 kOe. The nonlinear nature of $M(H)$ is, however, decreased with Ru substitution where $M(H)$ data for highest doped sample i.e., $x$ = 0.4 show almost linear behavior (Fig. 6a). The figure further shows magnetic moment decreases in this series. A close observation in $M(H)$ data reveals a narrow hysteresis in parent as well as in Ru-doped compounds. The observed coercive field $H_c$ is given in Table 1 which shows $H_c$ value increases as we introduce Ru in Y$_2$Ir$_2$O$_7$. Fig. 6b shows Arrott plot ($M^2$ vs $H/M$) of $M(H)$ data as shown in Fig. 6a. Arrott plot in general is a very helpful tool to understand the nature of magnetic state in a material.\cite{arrott} For instance, a straight line fitting in high field regime in Arrott plot gives a positive intercept for the materials having spontaneous magnetization like ferromagnetic compounds. In addition, value of the intercept gives an estimate of strength of spontaneous magnetization in the system. On contrary, a negative intercept in Arrott plot implies a non-ferromagnetic type of magnetic state. As seen in Fig. 6b, the intercept for all the materials in present series is negative which indicates low temperature magnetic state in Y$_2$Ir$_2$O$_7$ is not of ferromagnetic type, and further Ru substitution also does not induce ferromagnetism in this material.

\begin{figure}
	\centering
		\includegraphics[width=9cm]{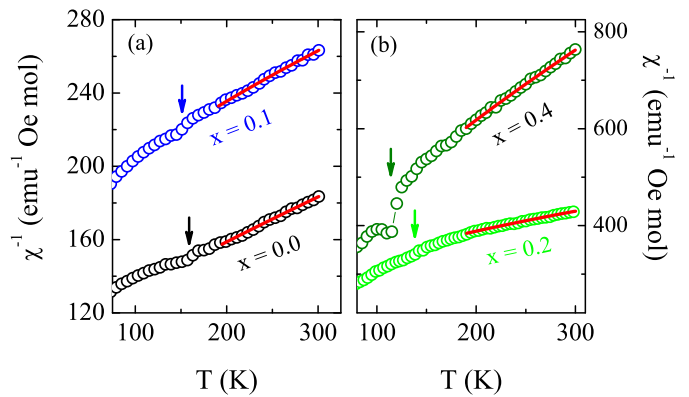}
	\caption{Temperature dependent inverse susceptibility ($\chi^{-1}$ = $(M/H)^{-1}$) as deduced from magnetization data in Fig. 4 are shown with (a) $x$ = 0.0, 0.1 and (b) $x$ = 0.2, 0.4 composition for Y$_2$Ir$_{2-x}$Ru$_x$O$_7$ series. The solid lines are due to fitting using Eq. 1. The arrows indicate the T$_{irr}$ of corresponding materials.}
	\label{fig:Fig5}
\end{figure}

\begin{table}[b]
\caption{\label{tab:table 1} Parameters obtained from fitting of the magnetization data by modified Curie-Weiss expression $\chi$ = $M/H$ = $\chi_0$ + $C/(T - \theta_P)$ for Y$_2$Ir$_{2-x}$Ru$_x$O$_7$.}
\begin{ruledtabular}
\begin{tabular}{cccccc}
$x$ & $C$ & $\theta_P$ & $\mu_{eff}$ & $f$ & $H_c$\\
&(emu K mol$^{-1}$) &(K) &($\mu_B/f.u.$) & &(Oe)\\ 
\hline
0.0 &2.83 &-331 &4.75 &2.07 &100\\
0.1 &1.39 &-312 &3.33 &2.04 &377\\
0.2 &0.69 &-282 &2.34 &2.01 &521\\
0.4 &0.43 &-127 &1.85 &1.06 &555\\
\end{tabular}
\end{ruledtabular}
\end{table}

Evolution of magnetic behavior in Y$_2$Ir$_{2-x}$Ru$_x$O$_7$ series where both the magnetic moment as well as T$_{irr}$ decreases (Figs. 4 and 6) with Ru is rather interesting. The 4$d^4$ electronic state of Ru$^{4+}$ gives moment 2$\mu_B$/Ru considering only spin contribution. On the other hand, Ir$^{4+}$ has 5$d^5$ electronic state. While all the $d$-electrons occupy low energy $t_{2g}$ level, the present SOC effect leads to further splitting of $t_{2g}$ level into an effective pseudospin, $j_{eff}$ = 1/2 doublet and $j_{eff}$ = 3/2 quartet where the doublet has higher in energy.\cite{kim} This situation gives $j_{eff}$ = 3/2 levels are fully-filled and $j_{eff}$ = 1/2 levels are half-filled. Therefore, in strong SOC dominated picture, Ir gives moment (g$_j$$j_{eff}\mu_B$) of value 0.33$\mu_B$/Ir.  The role of $U$ here is to further split the $j_{eff}$ = 1/2 level and open a Mott-like gap which makes these materials $j_{eff}$ = 1/2 insulators. Decreasing the effective moment in Y$_2$Ir$_{2-x}$Ru$_x$O$_7$, even with substitution of high moment Ru for low one Ir is quite surprising. Note, that Ru substitution in other class of iridates has also interesting effects. For instance, moment in Sr$_2$Ir$_{1-x}$Ru$_x$O$_4$ is seen to decrease with progressive Ru content.\cite{cava,yuan,calder} In case of Na$_2$Ir$_{1-x}$Ru$_x$O$_3$, the Ru weakens the long-range antiferromagnetic ordering and and induces spin-glass like state even with smallest level of substitution.\cite{kavita} Apart from strong SOC picture, recent theoretical studies have further considered the possible influence of non-cubic crystal field such as, trigonal crystal field on the magnetic and electronic properties in iridium oxides.\cite{yang,liu} Whether such effects apart from considering only strong SOC are responsible for present evolution of magnetic behavior needs to be investigated.        

\begin{figure}
	\centering
		\includegraphics[width=8.5cm]{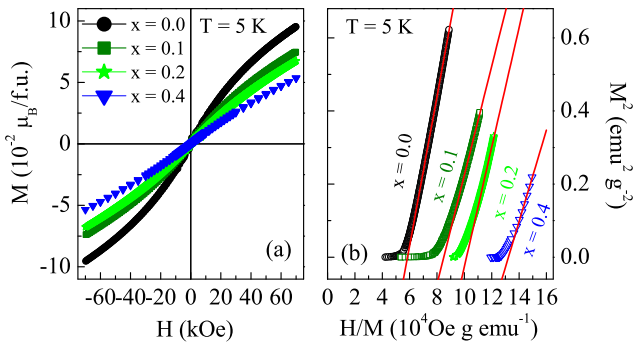}
	\caption{(a) Magnetic field dependent magnetization are shown for Y$_2$Ir$_{2-x}$Ru$_x$O$_7$ at 5 K. (b) The M(H) data are plotted in the form of Arrott plot ($M^2$ vs $H/M$) for the same materials.}
	\label{fig:Fig6}
\end{figure} 

As seen in Fig. 4, materials in Y$_2$Ir$_{2-x}$Ru$_x$O$_7$ series show considerable magnetic irreversibility between $M_{ZFC}$ and $M_{FC}$ below $T_{irr}$, and both the magnetization data exhibit steep rise at low temperature. A recent study has further shown reasonable magnetic relaxation and aging behavior at low temperature in Y$_2$Ir$_2$O$_7$.\cite{harish} These features which suggest nonequilibrium magnetic state is considered to be typical behavior of metamagnetic systems such as, spin-glass or superparamagnet as the magnetization for classical magnetic systems ideally do not change with time once temperature and filed is stabilized. We have examined the evolution of magnetic relaxation and aging behavior in present series. For the relaxation measurement, the sample has been cooled in zero magnetic field from room temperature to 10 K. Once temperature is stabilized, a time of 10$^2$ s is waited and then magnetic field of 1000 Oe is applied and subsequently magnetization has been measured as a function of time ($t$) for about 7200 s. Fig. 7a shows magnetization data after normalizing with magnetization value at $t$ = 0 i.e., with $M(0)$ as a function of time for representative $x$ = 0.0 and 0.2 samples. While the normalized $M(t)/M(0)$ continue to increase without sign of saturation even after 2 h for both the materials, it is seen in figure that rate of relaxation is comparatively higher in doped sample than the parent compound. This magnetic relaxation implies spin configuration at low temperature is not at equilibrium state as it tries to achieve low energy states with time crossing the barriers which separate the energy states. This data suggest Ru$^{4+}$ (4$d^4$) substitution at Ir$^{4+}$ (5$d^5$) site helps the spin system to relax at faster rate. 

\begin{figure}
	\centering
		\includegraphics[width=8.5cm]{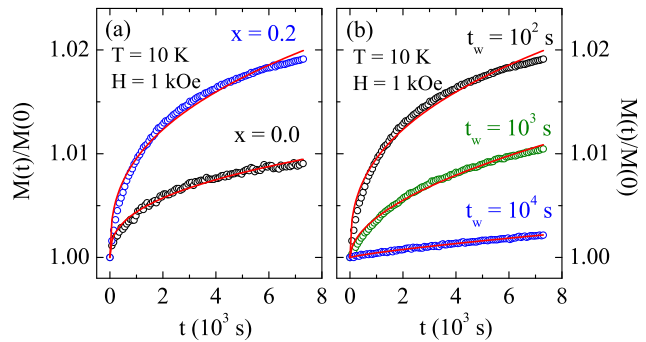}
	\caption{(a) The normalized magnetic moment $M(t)/M(0)$ as a function of time have been shown for $x$ = 0.0 and 0.2 with $t_w$ = 10$^2$ s. (b) The similar normalized moment have been shown for different wait time $t_w$ for Y$_2$Ir$_{1.6}$Ru$_0.4$O$_7$. The solid lines show representative fitting of data using Eq. 2.}
	\label{fig:Fig7}
\end{figure}

We have further examined the aging behavior in this series of materials. For the aging measurement, similarly the sample has been cooled in zero field from room temperature to 10 K. Once temperature is stabilized a wait time $t_w$ is given and after that a magnetic field of 1000 Oe is applied. Once field is applied, magnetization has been measured as a function of time for about 7200 s. It can be mentioned that in previous relaxation measurements we have used $t_w$ = 10$^2$ s and here in aging measurements we have varied $t_w$ with values 10$^2$, 10$^3$ and 10$^4$ s to see the effects of $t_w$. The aging behavior has already been observed for parent Y$_2$Ir$_2$O$_7$ material which shows considerable effect of $t_w$ on magnetic relaxation parameters.\cite{harish} Here in Fig. 7b we only show aging behavior of doped $x$ = 0.2 material by demonstrating normalized magnetization i.e., $M(t)/M(0)$ as a function of time for different $t_w$ values. Though the magnetization for all $t_w$ increase with time, it is evident in figure that system waits more at 10 K before field is applied shows less relaxation. This aging behavior is similar to that observed for parent compound (not shown) which suggests magnetic relaxation depends on the history how magnetic field is applied. This implies system ages during the wait time $t_w$ and tries to achieve equilibrium state which explains low relaxation with high $t_w$ value.

To understand this relaxation phenomena we have analyzed the $M(t)$ data with stretched exponential function as given below,\cite{chamberlin}

\begin{eqnarray}
	M(t) = M(0)\exp \left(\frac{t}{\tau}\right)^\beta
\end{eqnarray}
 
where $M(0)$ is magnetization value at $t$ = 0, $\tau$ is the characteristic relaxation time and $\beta$ is the stretching exponent ranging between 0 and 1. The solid lines in Fig. 7a and 7b are due to fitting of data with Eq. 2. It is evident in figure that fittings are reasonably good for all the data. The fitting parameters $\tau$ and $\beta$ are shown in Table II for both the samples. As seen in Table II, relaxation time $\tau$ decreases almost by an order after doping with Ru for all $t_w$ values. This decrease of $\tau$ and simultaneous increase of exponent $\beta$ (Table II) is in agreement with higher relaxation behavior for doped samples as evident in Fig. 7a. Table II further shows with increasing value of $t_w$, the nature of changes of $\tau$ and $\beta$ parameters, where $\tau$ decreases and $\beta$ increases, is same for both the samples. This clearly suggests that Ru$^{4+}$ (4$d^4$) substitution at Ir$^{4+}$ (5$d^5$) site helps the spin system to attain equilibrium state at faster rate, where the behavior is supported by reduced frustration parameter in doped sample (see in Table I).

\begin{table}
\caption{\label{tab:table 2} The characteristic relaxation time $\tau$ and the exponent $\beta$ as obtained from fitting of magnetic relaxation data with Eq. 2 (Fig. 7) are shown for different waiting time $t_w$ for $x$ = 0.0 and 0.2 sample of Y$_2$Ir$_{2-x}$Ru$_{x}$O$_7$ series.}
\begin{ruledtabular}
\begin{tabular}{cccc}
$t_w (s)$ &$x$ &$\tau$ (s) &$\beta$\\      
\hline
10$^2$  &0.0 &5.6 $\times$ 10$^9$ &0.33(1)\\
   &0.2 &3.1 $\times$ 10$^8$ &0.36(1)\\
\hline
10$^3$  &0.0 &1.78 $\times$ 10$^9$ &0.41(2)\\
  &0.2 &0.42 $\times$ 10$^8$ &0.52(2)\\
\hline
10$^4$  &0.0 &0.1 $\times$ 10$^9$ &0.82(8)\\
  &0.2 &0.16 $\times$ 10$^8$ &0.79(2)\\
\end{tabular}
\end{ruledtabular}
\end{table}

The Y$_2$Ir$_2$O$_7$ is known to show insulating property throughout the temperature range where the electrical resistivity ($\rho$) follows a power-law dependence on temperature.\cite{disseler,harish} The substitution of Ir$^{4+}$ (5$d^5$) with Ru$^{4+}$ (4$d^4$) basically amounts to hole doping in system which is likely to modify the electronic structure of material across the Fermi level. In addition, electronic correlation effect ($U$) as well as SOC are also expected to be tuned with this substitution. To understand this, we have studied the evolution of electrical resistivity in Y$_2$Ir$_{2-x}$Ru$_{x}$O$_7$ series. Fig. 8a shows temperature dependence of resistivity for this series. The figure shows with this Ru substitution (up to $x$ = 0.4), the system retains insulating but resistivity is substantially reduced. Given that another end member of this series i.e., Y$_2$Ru$_2$O$_7$ exhibits highly insulating state this decrease in resistivity appears to be interesting. As the nature of charge transport in parent compound obeys following power-law behavior

\begin{eqnarray}
	\rho = \rho_0 T^{-n}
\end{eqnarray}

where $n$ is power exponent, we have plotted $\rho(T)$ data in ln-ln scale in Fig. 8b. The straight lines in figure are due to fitting with Eq. 3. For $x$ = 0.0 material, power-law behavior (Eq. 3) is followed throughout the temperature range with exponent $n$ = 2.98. For the doped samples, we observe Eq. 3 could only be fitted in lower range of temperatures. Fig. 8b shows range of fitting extends up to 84, 65 and 72 K and we obtain exponent $n$ = 3.41, 4.01 and 1.96 for $x$ = 0.1, 0.2 and 0.4, respectively. This shrinking of fitting regime as well as variation of $n$ could be due to fast change in $\rho(T)$ as seen in Fig. 8a.

\begin{figure}
	\centering
		\includegraphics[width=8.5cm]{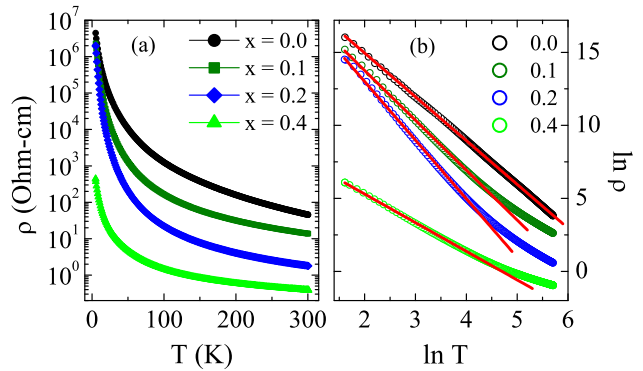}
	\caption{(a) Temperature dependent resistivity data of Y$_2$Ir$_{2-x}$Ru$_{x}$O$_7$ series are shown in semi-log plot. (b) The $\rho(T)$ data in (a) are plotted in $\ln-\ln$ plot. The solid lines are due to fitting with Eq. 3.}
	\label{fig:Fig8}
\end{figure}

The insulating state in various Ir-based oxides are of general interest which have been investigated using different substitutions. In case of Ru$^{4+}$ substitution in layered Sr$_2$IrO$_4$, resistivity is seen to decrease considerably where the another end member i.e., Sr$_2$RuO$_4$ is highly metallic.\cite{cava,maeno} Given that resistivity of Y$_2$Ru$_2$O$_7$ is very high,\cite{yoshii} this decrease in resistivity in Y$_2$Ir$_{2-x}$Ru$_x$O$_7$ is rather interesting. It remains fact that substitution of Ru$^{4+}$ in Y$_2$Ir$_2$O$_7$ will lead to depletion of effective electronic concentration in system (hole doping) and it will also tune both SOC and $U$ accordingly. Nonetheless, the influence of structural modification on electronic structure is an important factor which needs to be considered.\cite{yang} As seen in Figs. 2a and 2b that lattice parameter $a$ and trigonal crystal field represented by `$x$-position of O1 atom' decreases with Ru substitution, respectively. Further, Ir-O1 bond length decreases and Ir-O1-Ir bond angel increases as seen in Figs. 2c and 2d. The structural parameters in Figs. 2c and 2d imply that overlapping between Ir/Ru $d$-orbitals and O $p$-orbitals increases with Ru doping. This is likely to facilitate the transfer of charge carriers which would decrease the resistivity of material. However, any possible role of opposing tuning of SOC and $U$ realized through Ru (4$d$) substitution on electronic transport need to be investigated using theoretical calculations.

Further, it would be interesting to understand how the variations of SOC and $U$ realized through substitution of Ru in Y$_2$Ir$_2$O$_7$ contribute to the evolution of magnetic and electronic properties. The pyrochlore lattice with Heisenberg type nearest-neighbor interactions generally shows collinear AFM spin structures where frustration is very inherent feature. The additional consideration of DMI, mainly originating from present SOC, results in non-collinear type magnetic structures.\cite{elhajal} The decrease of SOC with Ru substitution will presumably drive the system toward more collinear type AFM spin structure, yet the Mossbauer spectroscopy experiments have shown another end compound i.e., Y$_2$Ru$_2$O$_7$ is still a non-collinear type AFM system.\cite{kmiec} Moreover, recent calculations have shown that with increasing $U$ the magnetic state in Y$_2$Ir$_2$O$_7$ moves from non-collinear AIAO state to AFM spin structure.\cite{shinaoka,Ishii} Therefore, the decreasing moment with $x$ in present study (see Figs. 4 and 6) may be a consequence of variation of both SOC and $U$. Similarly, decreasing resistivity with Ru substitution (Fig. 8) may also be contributed by variations of SOC and $U$ but the role of hole doping in modifying the Fermi Level as well as the structural modifications which are discussed in earlier sections need to be considered.

Even though, Ru$^{4+}$ is expected to modify the total electron numbers in system (hole doping), effects of SOC and $U$ variations on the magnetic and electronic properties is quite evident in our experimental data. We want to further add that initially it was believed that isoelectronic substitution could be achieved through doping with another 4$d$ element Rh$^{4+}$ for Ir$^{4+}$,\cite{lee} but recent experimental evidences from x-ray absorption spectroscopy (XAS) and angel resolved photoemission spectroscopy (ARPES) reveal that Rh rather adopts Rh$^{3+}$ charge state which amounts another case of hole doping.\cite{clancy,sohn,y-cao} Moreover, effective number of electrons estimated from optical conductivity data for Sr$_2$IrO$_4$ and Nd$_2$Ir$_2$O$_7$ materials show its evolution with Rh substitution level is not consistent.\cite{lee,ueda} In addition, if it is Rh$^{3+}$, this charge state will create equivalent amount of Ir$^{5+}$ ions which are nonmagnetic and will act for site dilution. This would further complicate the situation, making it difficult to understand the mere effects of variations of SOC and $U$. With this background, substitution of Ru$^{4+}$ appears to be a better mean to understand the effects of SOC and $U$ in iridate materials. Given that low temperature magnetic state in terms of long-range ordering in Y$_2$Ir$_2$O$_7$ is a controversial issue\cite{shapiro,disseler} and further it shows nonequilibrium  magnetic behavior,\cite{harish} the evolution of magnetic relaxation in Fig. 7 is quite intriguing. While the origin of magnetic relaxation can be many including glass-like dynamics and/or atomic displacement but based on these available bulk data it is difficult to make precise comment on the nature of magnetic state in Y$_2$Ir$_2$O$_7$. Nonetheless, our data definitely show nonequilibrium low temperature state in Y$_2$Ir$_2$O$_7$ as demonstrated through magnetic relaxation behavior which evolves with Ru substitution.            
   
\section{Conclusion}
In summary, we have investigated the structural, magnetic and electronic transport properties of pyrochlore iridate Y$_2$Ir$_{2-x}$Ru$_{x}$O$_7$ where the substitution of 4$d$ based Ru$^{4+}$ is aimed to tune the SOC and $U$ parameters. This substitution shows no structural phase transition, but the structural parameters are observed to evolve continuously with Ru doping. X-ray photoemission spectroscopy measurements indicate that Ru adopts Ru$^{4+}$ charge state and replaces the Ir$^{4+}$ as intensity of Ir$^{4+}$ decreases by $\sim$ 19.5\% which agrees with the nominal concentration of 20\% Ru in this material. The parent Y$_2$Ir$_2$O$_7$ shows magnetic irreversibility around 160 K which monotonically decreases with Ru doping. Further, magnetic moment also decreases with increasing Ru substitution. Magnetic relaxation measurement shows Y$_2$Ir$_2$O$_7$ and doped samples are in nonequilibrium magnetic state at low temperature. Interestingly, analysis of relaxation data shows relaxation rate increases with Ru doping. Resistivity data of Y$_2$Ir$_{2-x}$Ru$_{x}$O$_7$ show an insulating behavior throughout the temperature range for all the samples, however, resistivity decreases with Ru doping. The nature of charge conduction is found to follow the power-law behavior for all the materials.  

\section{Acknowledgment}   
We acknowledge UGC-DAE CSR, Indore for the measurements and providing User facility. We are thankful to Alok Banerjee and Rajeev Rawat for magnetic and resistivity measurements, and to Kranti Kumar and Sachin Kumar for the helps in measurements. HK acknowledges UGC, India for BSR fellowship.


\begin{thebibliography}{}
\bibitem{pesin} D. Pesin and L. Balents, Nature Phys \textbf{6}, 376 (2010).
\bibitem{william} W. Witczak-Krempa, G. Chen, Y. B. Kim, and L. Balents, Annual Rev. of Condens. Matter Physics. \textbf{5}, 57 (2014).
\bibitem{elhajal}  M. Elhajal, B. Canals, R. Sunyer and C. Lacroix, Phys. Rev. B \textbf{71}, 094420 (2005).
\bibitem{chen} G. Chen and M. Hermele, Phys. Rev. B \textbf{86}, 235129 (2012).
\bibitem{shapiro} M. C. Shapiro, S. C. Riggs, M. B. Stone, C. R. de la Cruz, S. Chi, A. A. Podlesnyak and I. R. Fisher, Phys. Rev. B \textbf{85}, 214434 (2012).
\bibitem{disseler} S. M. Disseler, C. Dhital, A. Amato, S. R. Giblin, C. de la Cruz, S. D. Wilson, and M. J. Graf, Phys. Rev. B \textbf{86}, 014428 (2012).
\bibitem{taira} N. Taira, M. Wakeshima and Y. Hinatsu, J. Phys.: Condens. Matter \textbf{13}, 5527 (2001).
\bibitem{fukazawa} H. Fukazawa and Y. Maeno, J. Phys. Soc. Jpn. 71, 2578 (2002).
\bibitem{soda} M. Soda, N. Aito, Y. Kurahashi, Y. Kobayashi, and M. Sato, Physica B \textbf{1071}, 329 (2003).
\bibitem{zhu} W. K. Zhu, M. Wang, B. Seradjeh, F. Yang, and S. X. Zhang, Phys. Rev. B \textbf{90}, 054419 (2014).
\bibitem{harish} H. Kumar and A. K. Pramanik, J. Magn. Magn. Mater \textbf{409} 20 (2016).
\bibitem{wan} X. Wan, A. M. Turner, A. Vishwanath, and S. Y. Savrasov, Phys. Rev. B \textbf{83}, 205101 (2011).
\bibitem{shinaoka} H. Shinaoka, S. Hoshino, M. Troyer, and P. Werner, Phys. Rev. Lett. \textbf{115} 156401 (2015).
\bibitem{Ishii} F. Ishii, Y. P. Mizuta, T. Kato, T. Ozaki, H. Weng, and S. Onoda, J. Phys. Soc. Jpn. \textbf{84}, 073703 (2015).
\bibitem{kalo} R. S. Singh, V. R. R. Medicherla, K. Maiti, and E. V. Sampathkumaran, Phys. Rev. B \textbf{77}, 201102, (2008).
\bibitem{yang} B-J. Yang and Y. B. Kim, Phys. Rev. B \textbf{82}, 085111 (2010). 
\bibitem{young} R. A. Young, A. Sakthivel, T. S. Moss and C. O. Paiva-Santos, \textsl{Users guide to program DBWS-9411, Atlanta: Georgia Institute of Technology} (1994).
\bibitem{xps} J. F. Moulder, W. F. Stickle, P. E. Sobol, and K. D. Bomben, \textit{Handbook of X-ray Photoelectron Spectroscopy, Perkin-Elmer Corporation, USA} (1992).
\bibitem{gingras} M. J. P. Gingras, C. V. Stager, N. P. Raju, B. D. Gaulin, and J. E. Greedan, Phys. Rev. Lett. \textbf{78}, 947 (1997).
\bibitem{yoshii} S. Yoshii and M. Sato, J. Phys. Soc. Jpn. 68, 3034 (1999).
\bibitem{gardner1} J. S. Gardner, S. R. Dunsiger, B. D. Gaulin, M. J. P. Gingras, J. E. Greedan, R. F. Kiefl, M. D. Lumsden, W. A. MacFarlane, N. P. Raju, J. E. Sonier, I. Swainson, and Z. Tun, Phys. Rev. Lett. \textbf{82}, 1012 (1999).
\bibitem{nakatsuji} S. Nakatsuji, Y. Machida, Y. Maeno, T. Tayama, T. Sakakibara, J. van Duijn, L. Balicas, J. N. Millican, R. T. Macaluso and J. Y. Chan, Phys. Rev. Lett. \textbf{96}, 087204 (2006).
\bibitem{bramwell} S. T. Bramwell, M. J. Harris, B. C. den Hertog, M. J. P. Gingras, J. S. Gardner, D. F. McMorrow, A. R. Wildes,
A. L. Cornelius, J. D. M. Champion, R. G. Melko, and T. Fennell, Phys. Rev. Lett. \textbf{87}, 047205 (2001).
\bibitem{fukazawa1} H. Fukazawa, R. G. Melko, R. Higashinaka, Y. Maeno, and M. J. P. Gingras, Phys. Rev. B \textbf{65}, 054410 (2002).
\bibitem{kmiec} R. Kmie$\acute{c}$, $\dot{Z}$. $\acute{S}$wiatkowska, J. Gurkul, M. Rams, A. Zarzycki and K. Tomala, Phys. Rev. B \textbf{74}, 104425 (2006).
\bibitem{Gardner} J.S. Gardner, B.D. Gaulin, S. H. Lee, C. Broholm, N.P. Raju, and J.E. Greedan, Phys. Rev. Lett. \textbf{83}, 211 (1999).
\bibitem{arrott} A. Arrott, Phys. Rev. \textbf{108}, 1394 (1957).
\bibitem{kim} B. J. Kim, Hosub Jin, S. J. Moon, J.-Y. Kim, B.-G. Park, C. S. Leem, Jaejun Yu, T.W. Noh, C. Kim, S.-J. Oh
J.-H. Park, V. Durairaj, G. Cao, and E. Rotenberg, Phys. Rev. Lett. \textbf{101}, 076402 (2008).
\bibitem{cava} R. J. Cava, B. Batlogg, K. Kiyono, H. Takagi, J. J. Krajewski, W. F. Peck, Jr., L. W. Rupp, Jr., and C. H. Chen, Phys. Rev. B \textbf{49}, 11890 (1994).
\bibitem{yuan} S. J. Yuan, S. Aswartham, J. Terzic, H. Zheng, H. D. Zhao, P. Schlottmann, and G. Cao, Phys. Rev. B \textbf{92}, 245103 (2015).
\bibitem{calder} S. Calder, J. W. Kim, G.-X. Cao, C. Cantoni, A. F. May, H. B. Cao, A. A. Aczel, M. Matsuda, Y. Choi, D. Haskel, B. C. Sales, D. Mandrus, M. D. Lumsden, and A. D. Christianson, Phys. Rev. B \textbf{92}, 165128 (2015).
\bibitem{kavita} K. Mehlawat, G. Sharma, and Y. Singh, Phys. Rev. B \textbf{92}, 134412 (2015).
\bibitem{liu} X. Liu, V. M. Katukuri, L. Hozoi, W-G. Yin, M. P. M. Dean, M. H. Upton, Jungho Kim, D. Casa, A. Said, T. Gog, T. F. Qi, G. Cao, A. M. Tsvelik, J. van den Brink, and J. P. Hill, Phys. Rev. Lett. \textbf{109}, 157401 (2012).
\bibitem{chamberlin} R. V. Chamberlin, G. Mozurkewich, and R. Orbach, Phys. Rev. Lett. \textbf{52}, 867 (1984).
\bibitem{maeno} Y. Maeno, T. Ando, Y. Mori, E. Ohmichi, S. Ikeda, S. NishiZaki and S. Nakatsuji, Phys. Rev. Lett. \textbf{81}, 3765 (1998).
\bibitem{lee} J. S. Lee, Y. Krockenberger, K. S. Takahashi, M. Kawasaki, and Y. Tokura, Phys. Rev. B \textbf{85}, 035101 (2012). 
\bibitem{clancy} J. P. Clancy, A. Lupascu, H. Gretarsson, Z. Islam Y. F. Hu, D. Casa, C. S. Nelson, S. C. LaMarra,
G. Cao, and Young-June Kim, Phys. Rev. B \textbf{89}, 054409 (2014).
\bibitem{sohn} C. H. Sohn, Deok-Yong Cho, C.-T. Kuo, L. J. Sandilands, T. F. Qi, G. Cao, and T. W. Noh, Sci. Rep. \textbf{6}, 23856 (2016).
\bibitem{y-cao} Y. Cao, Q. Wang, J. A. Waugh, T. J. Reber, H. Li, X. Zhou, S. Parham, S.-R. Park, N. C. Plumb, E. Rotenberg, A. Bostwick, J. D. Denlinger, T. Qi, M. A. Hermele, G. Cao, and Daniel S. Dessau, Nat. Commun. \textbf{7}, 11367 (2016).
\bibitem{ueda}  K. Ueda, J. Fujioka, Y. Takahashi, T. Suzuki, S. Ishiwata, Y. Taguchi, and Y. Tokura, Phys. Rev. Lett. \textbf{109}, 136402 (2012).  
\end{thebibliography}
\end{document}